\documentclass[aps,twocolumn,pra,superscriptaddress,amsmath,showpacs,tightenlines]{revtex4}
\usepackage{epsfig,graphicx,times}
\usepackage{amstext}
\usepackage{amsmath}            
\usepackage{amssymb}            
\usepackage{graphicx}           
\usepackage{latexsym}
\usepackage{bm}
\usepackage[colorlinks,citecolor=blue, linkcolor=blue,hyperindex,CJKbookmarks,dvipdfm]{hyperref}

\begin{document}

\title{Strongly correlated two-photon transport in a one-dimensional
waveguide coupled to a weakly nonlinear cavity}
\author{Xun-Wei Xu}
\affiliation{Beijing Computational Science Research Center, Beijing 100084, China}
\author{Yong Li}
\email{liyong@csrc.ac.cn}
\affiliation{Beijing Computational Science Research Center, Beijing 100084, China}
\affiliation{Synergetic Innovation Center of Quantum Information and Quantum Physics,
University of Science and Technology of China, Hefei, Anhui 230026, China}
\date{\today }

\begin{abstract}
We study the photon-photon correlation properties of two-photon transport in
a one-dimensional waveguide coupled to a nonlinear cavity via a real-space
approach. It is shown that the intrinsic dissipation of the nonlinear cavity
has an important effect upon the correlation of the transported photons.
More importantly, strongly correlated photons can be obtained in the transmitted
photons even when the nonlinear interaction strength is weak in the cavity. The strong photon-photon
correlation is induced by the Fano resonance involving destructive interference between the plane wave and
bound state for two-photon transport.
\end{abstract}

\maketitle


\section{Introduction}

Strong photon-photon interaction is one of the fundamental conditions for
the application of single photons in quantum information processing. Imamoglu \emph{et al}. proposed
to create strong photon-photon interaction by a high-finesse cavity
containing a low-density four-level atomic medium~\cite{ImamogluPRL97}. They
theoretically showed that strong antibunching effect can be observed in the transmitted photons when the Kerr
nonlinearity induced by the resonant atomic medium is strong. The underlying physics can be understood as the excitation of a
first photon blocking the transport of a second photon for the strong
nonlinearity in the cavity, which is referred as the photon blockade effect~\cite{ImamogluPRL97}.
Weakly driven cavity with strong nonlinearity is one of the systems for creating strongly interacting photons.
Beyond the traditional nonlinear optics, strong interactions between single
atoms and photons have already been demonstrated in the cavity quantum
electrodynamics systems~\cite{HoodPRL98,ReithmaierNat04,WallraffNat04,HennessyNat07}, and photon blockade
has been observed in the regime for strong atom-cavity coupling~\cite{BirnbaumNat05,DayanSci08,FaraonNP08,LangPRL11,HoffmanPRL11}.

Shen and Fan proposed another scheme to create strong
photon-photon interaction~\cite{ShenPRL07, ShenPRA07}. They showed that two-photon transport is strongly
correlated in one-dimensional (1D) waveguide coupled to a two-level system~%
\cite{ShenPRL07, ShenPRA07}. This strong correlation arises from the
interference between the reemitted and scattered waves in the 1D waveguide.
Photon-photon bound states appear due to the strong photon-photon
interaction and induce the photon blockade effect~\cite{ZhengPRA10, ZhengPRL11, ZhengPRA12}. Moreover, Roy proposed to realize
an optical diode at few-photon level by two- or multi-photon transport
in a 1D waveguide coupled asymmetrically to a two-level system~\cite%
{RoyPRB10}, or to generate a probe to detect atomic level structures by two-photon
scattering in a 1D waveguide coupled to different atomic system with similar
transition energies~\cite{RoyPRL11,RoyPRA13,RoySR13}. Strong photon-photon interaction
can also be created by two-photon transport in a 1D waveguide coupled to a
cavity containing a two-level atom~\cite{ShiPRA11,ShiPRA13}, two two-level atoms coupled via the Rydberg interaction~\cite{HuangPRA13}, a Kerr medium~\cite{LiaoPRA10}, or a four-level atom~\cite{YanPRA12}.

Recently, Liao and Law~\cite{LiaoPRA10} investigated the transport properties of two photons inside a
1D waveguide side-coupled to a single-mode nonlinear cavity based on the
Laplace transform method. They treated the nonlinear
cavity as a perfect one with no intrinsic dissipation, which is only applicable to the condition that the intrinsic
dissipation is much smaller than the coupling between the 1D waveguide and
single-mode cavity. In Refs.~\cite{ShenPRA09I,ShenPRA09II}, it has been revealed that the dissipation of the cavity
has a distinct effect on the single-photon transmission properties. To date, there exists little
literature on the subject that how the intrinsic dissipation of cavity
affects the transport properties of the two photons inside a 1D waveguide
side-coupled to a single-mode cavity with nonlinear medium. What's more, Liao and Law have shown
that strong photon-photon correlation between two transmitted or reflected
photons is based on strong nonlinearity~\cite{LiaoPRA10}. However,
it is difficult to obtain giant Kerr nonlinearity and low loss
simultaneously. So how to create strong photon-photon correlation with weak
nonlinearity should be an interesting subject.

In this paper, we will study the photon-photon correlation as two-photon
transport in 1D waveguide coupled to a nonlinear cavity by the real-space
approach~\cite{ShenPRL05,ShenPRA09I,ShenPRA09II} with taking the effect of the intrinsic loss of the nonlinear cavity
into account. It is shown that the correlation of the transport photons is significantly dependent on the intrinsic dissipation of the nonlinear cavity when the intrinsic dissipation is of the order of the coupling between the 1D waveguide and nonlinear cavity mode. More importantly, we find that the strong photon-photon correlation can be created even with weak nonlinearity in the cavity.

The paper is organized as
follows: In Sec.~II, we show the physical model for the transport of two
photons inside a 1D waveguide coupled to a nonlinear cavity. The photon-photon
correlation properties for the cavity with strong and weak nonlinearity are
investigated in Sec.~III and IV, respectively. Finally, we draw our
conclusions in Sec.~V.

\section{Physical model}

\begin{figure}[tbp]
\includegraphics[bb=100 218 495 553, width=6 cm, clip]{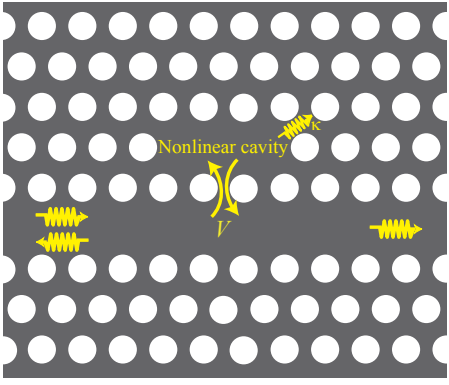}
\caption{(Color online) Schematic diagram of a 1D waveguide
coupled to a cavity with Kerr-type nonlinear medium. Photons injected into
the waveguide from the left
side are scattered by the nonlinear cavity, so the photons are
reflected or transmitted in the waveguide.}
\label{fig1}
\end{figure}

As shown in Fig.~\ref{fig1}, the system consists of a 1D waveguide (a row
defect waveguide) evanescently coupled to a nonlinear photonic crystal
cavity with coupling constant $V$~\cite{BoseOE11,BosePRL12,LiPRB14}. By
incorporating the excitation amplitudes of the reservoir and in a frame rotating at a frequency far away from the cutoff frequency of the dispersion for linearizing~\cite{ShenPRA09I,ShenPRA09II}, the effective
Hamiltonian of the system is~\cite{ShenPRL05} ($\hbar=1$):
\begin{eqnarray}
H &=&-iv_{c}\int dx\left[ c_{R}^{\dag }\left( x\right) \frac{\partial }{%
\partial x}c_{R}\left( x\right) -c_{L}^{\dag }\left( x\right) \frac{\partial
}{\partial x}c_{L}\left( x\right) \right]  \notag \\
&& +\int dxV\delta \left( x\right) \left[ c_{R}^{\dag }\left( x\right)
a+c_{L}^{\dag }\left( x\right) a+\mathrm{H.c.}\right]  \notag \\
&&+\left( \omega _{a}-i\frac{\kappa }{2}\right) a^{\dag }a+Ua^{\dag }a^{\dag
}aa,  \label{eq:1}
\end{eqnarray}%
where $c_{R}^{\dag }\left( x\right) $ [$c_{L}^{\dag }\left( x\right) $] is the
bosonic operator creating a right-going (left-going) photon with group
velocity $v_{c}$ at position $x$; $a$ ($a^{\dag }$) is the annihilation
(creation) operator of the cavity mode with frequency $\omega _{a}$; the cavity is filled with Kerr medium with nonlinear interaction strength $U$
and intrinsic dissipation rate $\kappa $. By employing $%
c_{e}^{\dag }\left( x\right) =\left[ c_{R}^{\dag }\left( x\right)
+c_{L}^{\dag }\left( -x\right) \right] /\sqrt{2}$, $c_{o}^{\dag }\left(
x\right) =\left[ c_{R}^{\dag }\left( x\right) -c_{L}^{\dag }\left( -x\right) %
\right] /\sqrt{2}$, the Hamiltonian is transformed into:%
\begin{eqnarray}
H &=&-iv_{c}\int dx\left[ c_{e}^{\dag }\left( x\right) \frac{\partial }{%
\partial x}c_{e}\left( x\right) +c_{o}^{\dag }\left( x\right) \frac{\partial
}{\partial x}c_{o}\left( x\right) \right]  \notag \\
&&+\int dx\overline{V}\delta \left( x\right) \left[ c_{e}^{\dag }\left(
x\right) a+c_{e}\left( x\right) a^{\dag }\right]  \notag \\
&&+\left( \omega _{a}-i\frac{\kappa }{2}\right) a^{\dag }a+Ua^{\dag }a^{\dag
}aa.  \label{eq:2}
\end{eqnarray}%
Here, the right- and left-going modes in the waveguide are transformed to
the even and odd modes \{$c_{e}^{\dag }\left( x\right) $, $c_{o}^{\dag
}\left( x\right) $\}, and the cavity mode only couples to the even mode with
effective coupling constant $\overline{V}=\sqrt{2}V$.

Assume that there are two photons injected into the waveguide from the left
side with momenta $k_{1}$ and $k_{2}$, respectively. The wave function for
the two photons before scattering (incoming state) is given by
\begin{equation}
\left\vert \Psi _{i}\right\rangle =\int \int dx_{1}dx_{2}
\phi_{k}\left(x_{1},x_{2}\right) \frac{1}{\sqrt{2}}c_{R}^{\dag }\left(
x_{1}\right) c_{R}^{\dag }\left( x_{2}\right) \left\vert \varnothing
\right\rangle ,  \label{eq:3}
\end{equation}%
where $\phi_{k}\left(x_{1},x_{2}\right)=\left(
e^{ik_{1}x_{1}+ik_{2}x_{2}}+e^{ik_{1}x_{2}+ik_{2}x_{1}}\right)/(2\sqrt{2
}\pi)$, $\left\vert \varnothing \right\rangle $ is the vacuum state of the
system. The general two-photon scattering state for two incident photons in
the right-going mode can be obtained by solving the Schr\"{o}dinger equation
with the incoming state [Eq.~(\ref{eq:3})] (see Appendix A).

The asymptotic two-photon outgoing scattering state in the spaces with the
right- and left-going modes is composed of two transmitted, two reflected, and
one transmitted plus one reflected photons as follows~\cite{ShenPRA07,RoyPRL11,RoyPRA13}
\begin{eqnarray}
&&\int \int dx_{1}dx_{2}t\left( x_{1},x_{2}\right) \frac{1}{\sqrt{2}}%
c_{R}^{\dag }\left( x_{1}\right) c_{R}^{\dag }\left( x_{2}\right) \left\vert
\varnothing ,g\right\rangle   \notag \\
&&+\int \int dx_{1}dx_{2}r\left( x_{1},x_{2}\right) \frac{1}{\sqrt{2}}%
c_{L}^{\dag }\left( x_{1}\right) c_{L}^{\dag }\left( x_{2}\right) \left\vert
\varnothing ,g\right\rangle   \notag \\
&&+\int \int dx_{1}dx_{2}rt\left( x_{1},x_{2}\right) c_{L}^{\dag }\left(
x_{1}\right) c_{R}^{\dag }\left( x_{2}\right) \left\vert \varnothing
,g\right\rangle ,
\end{eqnarray}%
where
\begin{eqnarray}
t\left( x_{1},x_{2}\right)  &=&t_{p}\left( x_{1},x_{2}\right) +t_{b}\left(
x_{1},x_{2}\right) ,  \label{eq:37} \\
r\left( x_{1},x_{2}\right)  &=&r_{p}\left( x_{1},x_{2}\right) +r_{b}\left(
x_{1},x_{2}\right) , \\
rt\left( x_{1},x_{2}\right)  &=&rt_{p}\left( x_{1},x_{2}\right)
+rt_{b}\left( x_{1},x_{2}\right) .
\end{eqnarray}%
$t_{p}\left( x_{1},x_{2}\right) $, $r_{p}\left( x_{1},x_{2}\right) $ and $%
rt_{p}\left( x_{1},x_{2}\right) $ correspond to the plane wave part,
\begin{eqnarray}
t_{p}\left( x_{1},x_{2}\right)  &=&\phi _{k}\left( x_{1},x_{2}\right)
\overline{t}_{k_{1}}\overline{t}_{k_{2}},  \label{eq:40} \\
r_{p}\left( x_{1},x_{2}\right)  &=&\phi _{k}\left( -x_{1},-x_{2}\right)
\overline{r}_{k1}\overline{r}_{k_{2}}, \\
rt_{p}\left( x_{1},x_{2}\right)  &=&\frac{1}{2\pi }e^{-i\frac{\omega }{2}%
\frac{x}{v_{c}}}\left( \overline{r}_{k_{2}}\overline{t}_{k1}e^{i2\Delta _{1}\frac{x_{c}}{v_{c}}} \right.\nonumber\\
&& \quad  \quad \quad  \quad \quad  \left. +\overline{t}_{k_{2}}\overline{r}_{k1}e^{-i2\Delta _{1}\frac{x_{c}}{v_{c}}}\right) ,
\end{eqnarray}%
where
\begin{eqnarray}
\overline{t}_{k_{i}} &=&\frac{v_{c}k_{i}-\omega _{a}+i\frac{\kappa }{2}}{%
v_{c}k_{i}-\omega _{a}+i\frac{\kappa +\Gamma}{2}}, \\
\overline{r}_{k_{i}} &=&\frac{-i\frac{\Gamma }{2}}{v_{c}k_{i}-\omega _{a}+i%
\frac{\kappa+ \Gamma}{2}}
\end{eqnarray}%
are the single-photon transmission and reflection amplitudes, $\Gamma =\overline{V}^{2}/v_{c}$, $\Delta _{i}=v_{c}k_{i}-\omega /2$ and $\Delta
_{a}=\omega _{a}-\omega /2$, $x_{c}=(x_{2}+x_{1})/2$ and $x=x_{2}-x_{1}$ are the center-of-mass and the relative coordinates, respectively. $t_{b}\left( x_{1},x_{2}\right) $, $r_{b}\left(
x_{1},x_{2}\right) $ and $rt_{b}\left( x_{1},x_{2}\right) $ are
contributions from the two-photon bound state,
\begin{eqnarray}
t_{b}\left( x_{1},x_{2}\right)  &=&\frac{1}{4}Be^{i\omega \frac{x_{c}}{v_{c}}%
}e^{\left[ -i2\Delta _{a}-\left( \kappa +\Gamma \right) \right] \frac{%
\left\vert x\right\vert }{2v_{c}}},  \label{eq:45} \\
r_{b}\left( x_{1},x_{2}\right)  &=&\frac{1}{4}Be^{-i\omega \frac{x_{c}}{v_{c}%
}}e^{\left[ -i2\Delta _{a}-\left( \kappa +\Gamma \right) \right] \frac{%
\left\vert x\right\vert }{2v_{c}}}, \\
rt_{b}\left( x_{1},x_{2}\right)  &=&\frac{1}{2\sqrt{2}}Be^{-i\frac{\omega }{2}%
\frac{x}{v_{c}}}e^{\left[ -i2\Delta _{a}-\left( \kappa +\Gamma \right) %
\right] \frac{\left\vert x_{c}\right\vert }{v_{c}}}.
\end{eqnarray}
where $B$ is dependent on the nonlinear interaction strength $U$ as given in Appendix A [Eq.~(\ref{eq:A30})].

Let us introduce two quantities to characterize the correlation between the two
transmitted photons and the correlation between the two reflected photons:
\begin{eqnarray}
\eta _{t} &=&\frac{\left\vert t\left( x_{1},x_{2}\right) \right\vert ^{2}}{%
\left\vert t_{p}\left( x_{1},x_{2}\right) \right\vert ^{2}},  \label{eq:48}
\\
\eta _{r} &=&\frac{\left\vert r\left( x_{1},x_{2}\right) \right\vert ^{2}}{%
\left\vert r_{p}\left( x_{1},x_{2}\right) \right\vert ^{2}}.
\end{eqnarray}
$\left\vert t\left( x_{1},x_{2}\right) \right\vert ^{2}$ and $\left\vert
r\left( x_{1},x_{2}\right) \right\vert ^{2}$ are the probabilities for
two-photon transmission and two-photon reflection, respectively; $\left\vert t_{p}\left(
x_{1},x_{2}\right) \right\vert ^{2}$ and $\left\vert r_{p}\left(
x_{1},x_{2}\right) \right\vert ^{2}$ are the probabilities of two photons
being transmitted and reflected by the system without photon-photon
interaction because of $B=0$ for $U=0$. So if two photons are transmitted (reflected) independently, we have $\eta _{t}=1$ ($\eta _{r}=1$); if $\eta _{t}\neq 1$ or $\eta _{r}\neq 1$, then we have
correlated two-photon state. As $\eta _{t}<1$ ($\eta _{r}<1$), effective
repulsive interaction between the two transmitted (reflected) photons is induced, causing suppression
on two-photon transmission (reflection), then photon blockade emerges; on the
contrary, as $\eta _{t}>1$ ($\eta _{r}>1$), there is effective attractive
interaction between the two transmitted (reflected) photons, causing enhanced two-photon
transmission (reflection), and photon-induced tunneling appears.

\section{Strongly nonlinear regime}

\begin{figure}[tbp]
\includegraphics[bb=67 255 515 595, width=8.5 cm, clip]{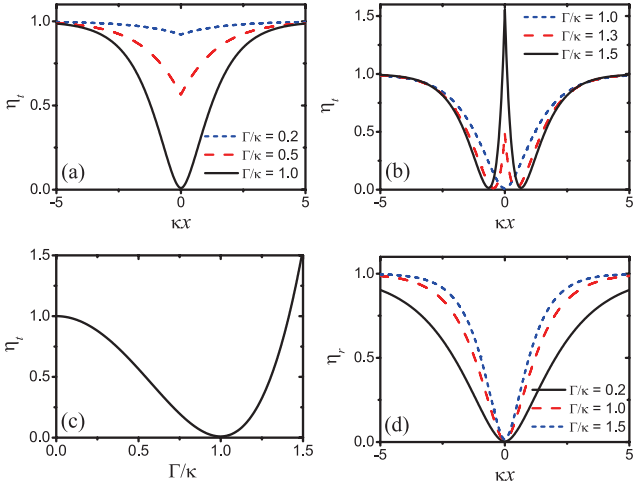}
\caption{(Color online) (a) and (b) $\protect\eta_{t}$ plotted as a function of
relative coordinate $\protect\kappa x$ for different values of the coupling constant $%
\Gamma / \protect\kappa$; (c) $\protect\eta_{t}$ plotted as a function of $%
\Gamma/\protect\kappa$ for $x=0$; (d) $\protect\eta_{r}$ plotted as a
function of relative coordinate $\protect\kappa x$ for different values of $\Gamma/\protect\kappa$. The other parameters are $%
\Delta_{1}=\Delta_{2}=\Delta_{a}=0$, $U/\protect\kappa=10$.}
\label{fig2}
\end{figure}

In this section, we will study the effect of the intrinsic
dissipation of the cavity on the two-photon transport properties with a strongly Kerr nonlinear medium in the cavity ($U>\kappa$). We assume that the two photons
are both resonant with the cavity mode, i.e. $\Delta_{1}=\Delta_{2}=%
\Delta_{a}=0$. In Figs.~\ref{fig2}(a) and (b), we show $\eta _{t}$ as
a function of relative coordinate $\kappa x$ for different values of the coupling parameter $%
\Gamma /\kappa$. At the point $x=0$, $\eta _{t}$ decreases from one to nearly zero as
$\Gamma/\kappa $ increases from zero to one, then increases
rapidly by the further increase of $\Gamma /\kappa$ when $\Gamma /\kappa >1$, as shown in Fig.~\ref%
{fig2}(c). Similar effect was also reported in Ref.~\cite{ZhengPRA10} for a
coherent sate transport in a 1D waveguide coupled to a single two-level
system. $\eta _{t}\approx 0$ at the point of $\Gamma /\kappa =1$ shows
that the two transmitted photons are strongly repulsive after transmitting; $\eta _{t}>1$
in the regime of $\Gamma /\kappa >1.4$ implies that the two transmitted photons are
attractive after transmitting. So we can make the transmitted photons from
repulsive to attractive by changing the coupling parameter $\Gamma /\kappa $.
$\eta _{r}$ as a function of relative coordinate $\kappa x$ for different values of the coupling parameter $\Gamma /\kappa$ is shown in Fig.~\ref{fig2}(d).
Around the point $x=0$, $\eta_{r}\approx 0$, the reflected photons are strongly repulsive and the window for $\eta _{r}<0$ becomes narrower with the increase of $\Gamma /\kappa$.

\begin{figure}[tbp]
\includegraphics[bb=81 408 508 567, width=8.5 cm, clip]{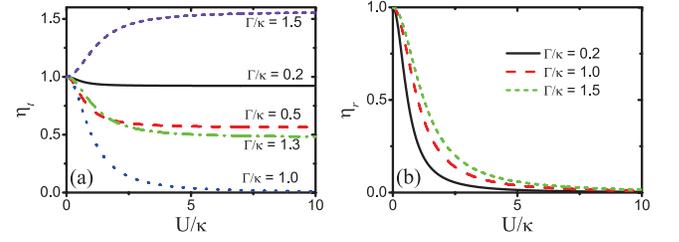}
\caption{(Color online) (a) $\protect\eta_{t}$ and (b) $\protect\eta_{r}$
plotted as functions of $U/\protect\kappa$ for different values of $\Gamma/%
\protect\kappa$ with $x=0$ and $\Delta_{1}=\Delta_{2}=\Delta_{a}=0$.}
\label{fig3}
\end{figure}

In Fig.~\ref{fig3}, we show $\eta _{t}$ and $\eta _{r}$\ as functions of
the nonlinear interaction strength $U/\kappa $ for different values of the coupling parameter $
\Gamma/\kappa $. For the transmitted photons, $\eta _{t}$ increases or decreases monotonously with the increase of $U$ depending on the value of $
\Gamma /\kappa$ and reaches the saturation point when $U\gg\kappa$. If we
want to obtain strongly repulsive transmitted photons, we should set $\Gamma
/\kappa =1$; on the contrary, if we need strongly attractive transmitted
photons, we can set $\Gamma /\kappa \gg 1$. For the reflected photons, $\eta
_{r}$ decreases monotonously as the increase of $U$ and is close to zero when $
U\gg\kappa$ and decreases more slowly with the increase of $\Gamma/\kappa$. That is
to say the reflected photons represent strongly repulsive interaction in the strongly
nonlinear interaction condition ($U\gg\kappa, \Gamma$) for $\Delta_{1}=\Delta_{2}=\Delta_{a}=0$.

\section{Weakly nonlinear regime}

\begin{figure}[tbp]
\includegraphics[bb=89 450 493 608, width=8.5 cm, clip]{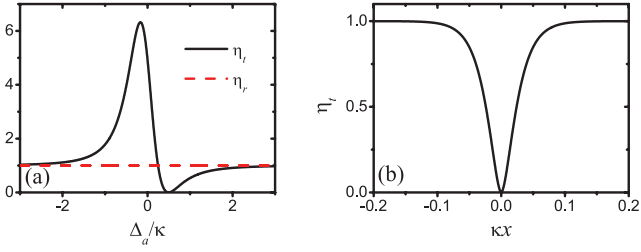}
\caption{(Color online) (a) $\eta_{t}$ and $\eta_{r}$ plotted as
functions of the detuning $\Delta_{a}/\protect\kappa$ for $x=0$, (b) $\protect%
\eta_{t}$ as a function of relative coordinate $\protect\kappa x$ for $%
\Delta_{a}=\protect\kappa/2$. The other parameters are $\Gamma=100 \protect%
\kappa$, $U=\protect\kappa/100$ and $\Delta_{1}=\Delta_{2}=0$.}
\label{fig4}
\end{figure}

Next, we will investigate whether strong photon-photon
correlation can be created with weak nonlinearity in the cavity ($U<\kappa$). $\eta _{t}$
and $\eta _{r}$ are plotted as functions of the detuning $\Delta_{a}/\kappa $
in Fig.~\ref{fig4}(a) for $\Delta_{1}=\Delta_{2}=0$, $\Gamma =100\kappa $,
and $U=\kappa/100$. We can see that there is an optimal point for the
transmitted photons exhibiting strongly repulsive interaction ($\eta _{t}\approx 0$) at $%
\Delta_{a} = \kappa/2$, while $\eta_{r}\approx 1$. The result shows that as
the coupling between the cavity mode and 1D waveguide is strong ($%
\Gamma=100\kappa $), the transmitted photons can exhibit strongly repulsive interaction
even when the nonlinear interaction in the cavity is weak ($U=\kappa /100$).
$\eta _{t}$ as a function of relative coordinate $\kappa x$ is shown in Fig.~%
\ref{fig4}(b). The spacescale of the repulsive interaction is much smaller than that of the cases in Figs.~\ref{fig2}(b) and (d).

\begin{figure}[tbp]
\includegraphics[bb=68 263 505 572, width=4.3 cm, clip]{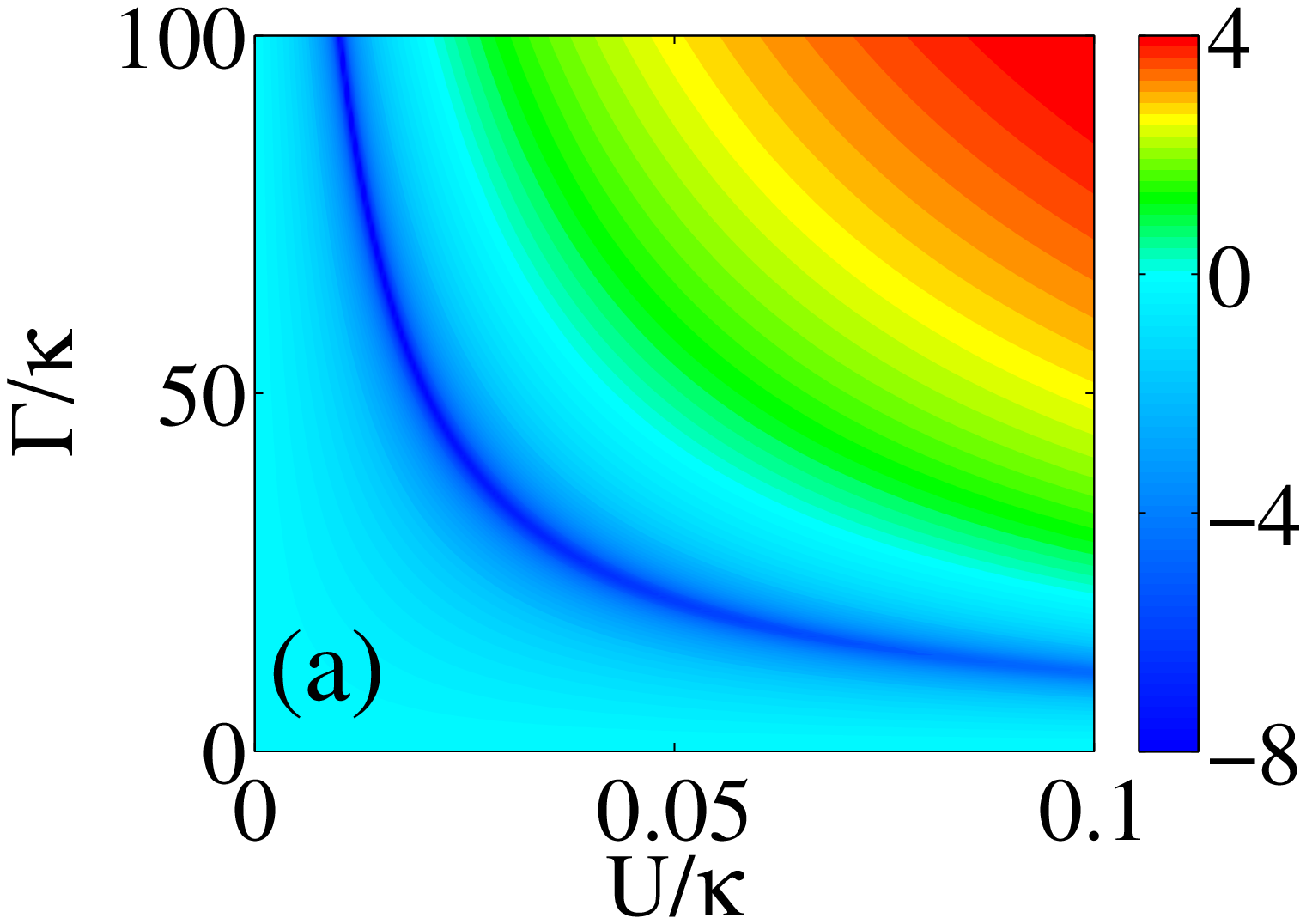} %
\includegraphics[bb=23 13 386 280, width=4.2 cm, clip]{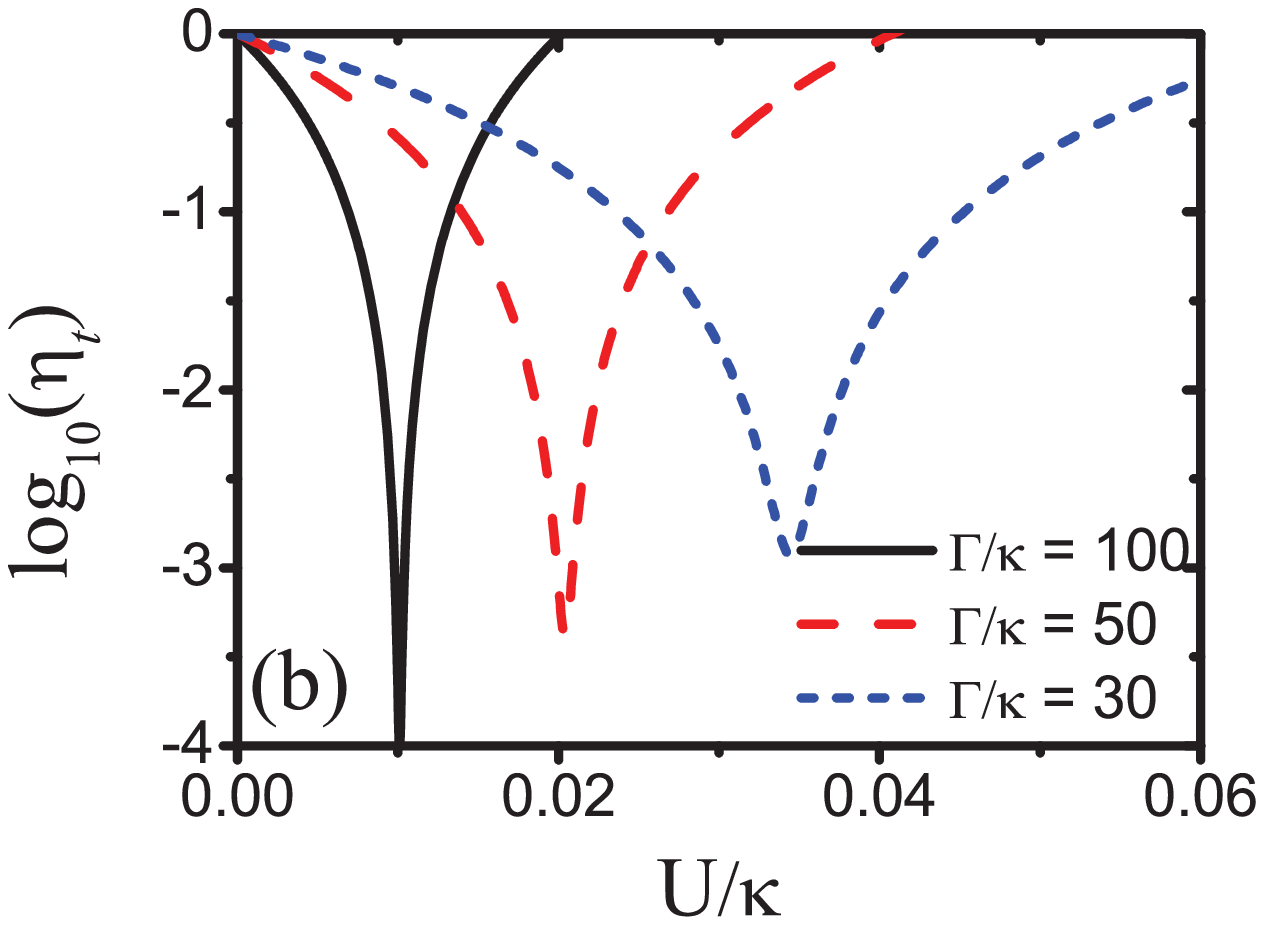}
\caption{(Color online) (a) $\log_{10}(\protect\eta_{t})$ plotted as
a function of $U/\protect\kappa$ and $\Gamma/\protect\kappa$. (b) $\log_{10}(%
\protect\eta_{t})$ plotted as a function of $U/\protect\kappa$ for different
values of $\Gamma/\protect\kappa$. The detunings $\Delta_{a}=\protect\kappa%
/2 $, $\Delta_{1}=\Delta_{2}=0$, and $x=0$.}
\label{fig5}
\end{figure}

To figure out the optimal conditions for obtaining the strongly repulsive
photon-photon interaction in the weakly nonlinear condition, a two-dimensional
plot of $\log_{10}(\eta_{t})$ as a function of the nonlinear interaction strength
$U/\kappa $ and coupling parameter $\Gamma/\kappa$ for $\Delta_{a} =\kappa/2
$ is shown in Fig.~\ref{fig5}(a). With increasing $\Gamma/\kappa$, the value
of $U/\kappa $ for getting the strongly repulsive interaction (dark blue
regime in the figure) descends gradually. From the cuts of the color scale
plot shown in Fig.~\ref{fig5}(b), for $\Delta_{a} =\kappa/2$, the optimal
strongly repulsive interaction occurs along the curve $\Gamma/\kappa \approx
\kappa/ U$.

In order to understand the origin of the strongly repulsive interaction, we will derive the optimal conditions
analytically after doing some approximations. From Eqs.~(\ref{eq:37}) and (%
\ref{eq:40}), as $x_{1}=x_{2}=x_{c}$ and $\Delta _{1}=\Delta _{2}=0$, we have%
\begin{eqnarray}
t\left( x_{1},x_{2}\right) &=&\frac{\sqrt{2}}{2\pi }e^{i\omega \frac{x_{c}}{%
v_{c}}}\overline{t}_{k_{1}}\overline{t}_{k_{2}}\chi, \\
t_{p}\left( x_{1},x_{2}\right) &=&\frac{\sqrt{2}}{2\pi }e^{i\omega \frac{%
x_{c}}{v_{c}}}\overline{t}_{k_{1}}\overline{t}_{k_{2}}
\end{eqnarray}
with
\begin{equation}
\chi = 1+\left( \frac{\Gamma }{2}\right) ^{2}\frac{U}{\left( \Delta _{a}+U-i%
\frac{\kappa +\Gamma }{2}\right) \left( \Delta _{a}-i\frac{\kappa }{2}%
\right) ^{2}}.
\end{equation}
In the condition that $\Gamma \gg \Delta _{a}\sim \kappa \gg U$, $\eta _{t}$ can be written in a Fano-like line shape approximately,
\begin{equation}
\eta _{t}\approx \frac{\epsilon ^{2}+\left( 1+q\right) ^{2}}{1+\epsilon ^{2}}
\label{eq:52}
\end{equation}%
with the parameters
\begin{eqnarray}
q &=&\frac{U\Gamma ^{2}}{2\Delta _{a}\left( 2\Delta _{a}^{2}-\kappa \Gamma
\right) }, \\
\epsilon &=&\frac{\Gamma \left( 4\Delta _{a}^{2}-\kappa ^{2}\right) }{%
4\Delta _{a}\left( 2\Delta _{a}^{2}-\kappa \Gamma \right) }.
\end{eqnarray}%
Eq.~(\ref{eq:52}) suggests that there is one minimum in the Fano-like
profile at
\begin{equation}
q=1,\quad \epsilon =0,
\end{equation}%
resulting in the optimal conditions for strongly repulsive interaction,%
\begin{equation}
\Delta _{a}=\frac{\kappa }{2},\quad \frac{U}{\kappa }=\frac{\kappa }{\Gamma }%
.  \label{eq:56}
\end{equation}%
After substituting the optimal conditions [Eq.~(\ref{eq:56})] back into
Eq.~(\ref{eq:48}), we have%
\begin{equation}
\eta _{t}\approx \left( \frac{U}{\kappa }\right) ^{2}.  \label{eq:57}
\end{equation}%
These analytical expressions [Eqs.~(\ref{eq:56}) and (\ref{eq:57})] agree well with the numerical results shown in Fig.~\ref{fig5}.

The strongly repulsive photon-photon interaction for the transmitted photons
comes from the destructive interference effect between different paths for
two-photon transmitting in the coupling system with Fano resonances~\cite%
{MiroshnichenkoRMP10}. The two photons can pass independently through the 1D waveguide coupled to a nonlinear cavity
as plan waves given by $t_{p}\left( x_{1},x_{2}\right)$ or instead
pass through the system together with a bound state given by $t_{b}\left( x_{1},x_{2}\right)$, which
is dependent on the strength of the nonlinearity. As $\Gamma \gg \kappa $,
the two photons are mainly resonantly reflected by the cavity in the regime $%
|\Delta _{a}|<V$, i.e. $\left\vert r_{p}\left( x_{1},x_{2}\right)
\right\vert ^{2}\approx 1$ and $\left\vert t_{p}\left( x_{1},x_{2}\right)
\right\vert ^{2}\approx 0$, so that the nonlinear interaction strength $U$
[in $t_{b}\left( x_{1},x_{2}\right) $] required for quantum destructive
interference can be much smaller than the intrinsic dissipation rate $\kappa $.

\section{Conclusions}

In conclusion, we have studied the transport properties of two photons in a
1D waveguide evanescently coupled to a nonlinear cavity. We have shown that
the correlation of the transport photons is strongly dependent on the intrinsic dissipation of the nonlinear cavity, and this provides us an effective way to control the
correlation properties for the transmitted two photons. What's more, due
to the Fano resonances involving destructive interference effect between the
plane wave and bound state for two-photon transmission, the transmitted two photons exhibit strongly repulsive interaction in the optimal conditions even with weak
nonlinearity in the cavity.

It is worth noting that the strong photon antibunching in the presence of weak
nonlinearity has also been found in two coupled nonlinear cavities
(nonlinear photonic molecule) theoretically~\cite{LiewPRL10,BambaPRA11,XuarXiv14a,XuarXiv14b}. The main difference is
that the strong photon antibunching was attributed to the ordinary destructive quantum
interference effect in the nonlinear photonic molecule~\cite{BambaPRA11}.
In addition, various nonlinear optical systems are proposed to achieve strong photon
blockade with weak nonlinearity, such as bimodal optical cavity with a
quantum dot~\cite{MajumdarPRL12,ZhangPRA14}, coupled optomechanical systems~\cite%
{XuJPB13,SavonaARX13}, or coupled single-mode cavities with second- or
third-order nonlinearity~\cite{FerrettiNJP13,FlayacPRA13,GeraceARX14}. Similarly, if a 1D waveguide is side-coupled to a cavity with a quantum dot, a second- or third-order nonlinear medium, or a mechanical resonator~\cite{LiaoPRA12,LiaoPRA13,RenPRA13,JiaPRA13} coupled to the cavity mode, the strong photon-photon interaction may also be obtained in the transmitted photons even when the effective nonlinearity in the cavity is weak.

\vskip 2pc \leftline{\bf Acknowledgement}

We thank Y. L. Liu, L. Ge, Q. Zheng, and Y. Yao for fruitful discussions.
This work is supported by the Postdoctoral Science Foundation of China (under Grant No. 2014M550019), the NSFC (under Grant No. 11174027), and the National 973 program (under Grant No. 2012CB922104 and No. 2014CB921402).

\appendix

\section{Two-photon scattering by nonlinear cavity}

\label{APP}

In this appendix we provide a derivation of the scattering states
for two incident photons in the right-going mode. Decomposing the right-going
mode to the even and odd modes by $c_{R}^{\dag }\left( x\right) =\left[
c_{e}^{\dag }\left( x\right) +c_{o}^{\dag }\left( x\right) \right] /\sqrt{2}$%
, the incoming state [Eq.~(\ref{eq:3})] can be rewritten as
\begin{widetext}
\begin{eqnarray}
\left\vert \Psi _{i}\right\rangle &=&\frac{1}{2}\int \int dx_{1}dx_{2}\frac{1%
}{2\pi \sqrt{2}}\left(
e^{ik_{1}x_{1}+ik_{2}x_{2}}+e^{ik_{1}x_{2}+ik_{2}x_{1}}\right) \frac{1%
}{\sqrt{2}}c_{e}^{\dag }\left( x_{1}\right) c_{e}^{\dag }\left( x_{2}\right)
\left\vert \varnothing \right\rangle  \notag \\
&&+\frac{1}{2}\int \int dx_{1}dx_{2}\frac{1}{2\pi \sqrt{2}}\left(
e^{ik_{1}x_{1}+ik_{2}x_{2}}+e^{ik_{1}x_{2}+ik_{2}x_{1}}\right) \frac{1%
}{\sqrt{2}}c_{o}^{\dag }\left( x_{1}\right) c_{o}^{\dag }\left( x_{2}\right)
\left\vert \varnothing \right\rangle  \notag \\
&&+\frac{1}{\sqrt{2}}\int \int dx_{1}dx_{2}\frac{1}{2\pi \sqrt{2}}\left(
e^{ik_{1}x_{1}+ik_{2}x_{2}}+e^{ik_{1}x_{2}+ik_{2}x_{1}}\right)
c_{e}^{\dag }\left( x_{1}\right) c_{o}^{\dag }\left( x_{2}\right) \left\vert
\varnothing \right\rangle .  \label{eq:4}
\end{eqnarray}%
We will determine the scattering state in the spaces spanned by the even and
odd modes.

The general two-photon scattering state of the system in the spaces with
even and odd modes takes the following form:
\begin{equation}
\left\vert \Psi \right\rangle =\frac{1}{2}\left\vert \Psi _{ee}\right\rangle
+\frac{1}{2\sqrt{2}}\left\vert \Psi _{oe}\right\rangle +\frac{1}{2\sqrt{2}}%
\left\vert \Psi _{eo}\right\rangle +\frac{1}{2}\left\vert \Psi
_{oo}\right\rangle ,  \label{eq:5}
\end{equation}%
with
\begin{eqnarray}
\left\vert \Psi _{ee}\right\rangle &=&\int \int dx_{1}dx_{2}\phi _{ee}\left(
x_{1},x_{2}\right) \frac{1}{\sqrt{2}}c_{e}^{\dag }\left( x_{1}\right)
c_{e}^{\dag }\left( x_{2}\right) \left\vert \varnothing \right\rangle +\int
dx\phi _{ae}\left( x\right) c_{e}^{\dag }\left( x\right) a^{\dag }\left\vert
\varnothing \right\rangle  \notag \\
&&+\phi _{aa}\frac{1}{\sqrt{2}}a^{\dag }a^{\dag }\left\vert \varnothing
\right\rangle ,  \label{eq:6} \\
\left\vert \Psi _{oe}\right\rangle &=&\int \int dx_{1}dx_{2}\phi _{oe}\left(
x_{1},x_{2}\right) c_{o}^{\dag }\left( x_{1}\right) c_{e}^{\dag }\left(
x_{2}\right) \left\vert \varnothing \right\rangle +\int dx\phi _{oa}\left(
x\right) c_{o}^{\dag }\left( x\right) a^{\dag }\left\vert \varnothing
\right\rangle , \\
\left\vert \Psi _{eo}\right\rangle &=&\int \int dx_{1}dx_{2}\phi _{eo}\left(
x_{1},x_{2}\right) c_{e}^{\dag }\left( x_{1}\right) c_{o}^{\dag }\left(
x_{2}\right) \left\vert \varnothing \right\rangle +\int dx\phi _{ao}\left(
x\right) c_{o}^{\dag }\left( x\right) a^{\dag }\left\vert \varnothing
\right\rangle , \\
\left\vert \Psi _{oo}\right\rangle &=&\int \int dx_{1}dx_{2}\phi _{oo}\left(
x_{1},x_{2}\right) \frac{1}{\sqrt{2}}c_{o}^{\dag }\left( x_{1}\right)
c_{o}^{\dag }\left( x_{2}\right) \left\vert \varnothing \right\rangle ,
\end{eqnarray}%
where $\phi _{ij}$ $(i,j=e,o,a)$ is the amplitude of the two photons with
one photon in mode $i$ and the other in mode $j$; subscript $e$ ($o$)
stands for even (odd) mode and subscript $a$ stands for cavity mode. In
order to satisfy the statistical property of photons, the amplitudes satisfy
the relations: $\phi _{ee}\left( x_{1},x_{2}\right) =\phi _{ee}\left(
x_{2},x_{1}\right) $, $\phi _{oo}\left( x_{1},x_{2}\right) =\phi _{oo}\left(
x_{2},x_{1}\right) $, $\phi _{eo}\left( x_{1},x_{2}\right) =\phi _{oe}\left(
x_{2},x_{1}\right) $, and $\phi _{oa}\left( x\right) =\phi _{ao}\left(
x\right) $.

In this paper, we concentrate on the two-photon transport of the frequency $\omega
=v_{c}k_{1}+v_{c}k_{2}$. In steady state, according to the time-independent Schr\"{o}dinger's equation, $H\left\vert \Psi
\right\rangle =\omega \left\vert \Psi \right\rangle $, we obtain the
following linear equations for the amplitudes in the scattering state:
\begin{eqnarray}
\left( -iv_{c}\frac{\partial }{\partial x_{1}}-iv_{c}\frac{\partial }{%
\partial x_{2}}-\omega \right) \phi _{ee}\left( x_{1},x_{2}\right) +\frac{%
\overline{V}}{\sqrt{2}}\left[ \delta \left( x_{1}\right) \phi _{ae}\left(
x_{2}\right) +\delta \left( x_{2}\right) \phi _{ae}\left( x_{1}\right) %
\right] &=&0, \\
\left( -iv_{c}\frac{\partial }{\partial x}+\omega _{a}-\omega -i\frac{\kappa
}{2}\right) \phi _{ae}\left( x\right) +\sqrt{2}\overline{V}\delta \left(
x\right) \phi _{aa}+\frac{\overline{V}}{\sqrt{2}}\left[ \phi _{ee}\left(
0,x\right) +\phi _{ee}\left( x,0\right) \right] &=&0, \\
\left( 2\omega _{a}-\omega +2U-i\kappa \right) \phi _{aa}+\sqrt{2}\overline{V%
}\phi _{ae}\left( 0\right) &=&0, \\
\left( -iv_{c}\frac{\partial }{\partial x_{1}}-iv_{c}\frac{\partial }{%
\partial x_{2}}-\omega \right) \phi _{oe}\left( x_{1},x_{2}\right) +%
\overline{V}\delta \left( x_{2}\right) \phi _{oa}\left( x_{1}\right) &=&0, \\
\left( -iv_{c}\frac{\partial }{\partial x}+\omega _{a}-\omega -i\frac{\kappa
}{2}\right) \phi _{oa}\left( x\right) +\overline{V}\phi _{oe}\left(
x,0\right) &=&0, \\
\left( -iv_{c}\frac{\partial }{\partial x_{1}}-iv_{c}\frac{\partial }{%
\partial x_{2}}-\omega \right) \phi _{oo}\left( x_{1},x_{2}\right) &=&0.
\end{eqnarray}%
We use $\phi _{ee}\left( x,0\right) =\left[ \phi _{ee}\left( x,0^{+}\right)
+\phi _{ee}\left( x,0^{-}\right) \right] /2$, $\phi _{ae}\left( 0\right) =%
\left[ \phi _{ae}\left( 0^{+}\right) +\phi _{ae}\left( 0^{-}\right) \right]
/2$, $\phi _{oe}\left( x,0\right) =\left[ \phi _{oe}\left( x,0^{+}\right)
+\phi _{oe}\left( x,0^{-}\right) \right] /2$ for the discontinuous points.
Solving the linear equations with the initial conditions [incoming state
given in Eq.~(\ref{eq:4})] and the discontinuity relations,
\begin{eqnarray}
\phi _{ee}\left( 0^{+},x\right) -\phi _{ee}\left( 0^{-},x\right) &=&\frac{%
\overline{V}}{iv_{c}\sqrt{2}}\phi _{ae}\left( x\right) , \\
\phi _{ee}\left( x,0^{+}\right) -\phi _{ee}\left( x,0^{-}\right) &=&\frac{%
\overline{V}}{iv_{c}\sqrt{2}}\phi _{ae}\left( x\right) , \\
\phi _{ae}\left( 0^{+}\right) -\phi _{ae}\left( 0^{-}\right) &=&\frac{\sqrt{2%
}\overline{V}}{iv_{c}}\phi _{aa}, \\
\phi _{oe}\left( x_{1},0^{+}\right) -\phi _{oe}\left( x_{1},0^{-}\right) &=&%
\frac{\overline{V}}{iv_{c}}\phi _{oa}\left( x_{1}\right) , \\
\phi _{oe}\left( 0^{+},x_{2}\right) &=&\phi _{oe}\left( 0^{-},x_{2}\right) ,
\\
\phi _{oa}\left( 0^{+}\right) &=&\phi _{oa}\left( 0^{-}\right) ,
\end{eqnarray}%
we find the amplitudes for the two-photon scattering state,%
\begin{eqnarray}
\phi _{ee}\left( x_{1},x_{2}\right) &=&\frac{1}{\sqrt{2}}\left[ \phi
_{e,k_{1}}\left( x_{1}\right) \phi _{e,k_{2}}\left( x_{2}\right) +\phi
_{e,k_{1}}\left( x_{2}\right) \phi _{e,k_{2}}\left( x_{1}\right) \right]
\notag \\
&&+\left[ \theta \left( x_{2}-x_{1}\right) \theta \left( x_{1}\right)
Be^{i\omega \frac{x_{c}}{v_{c}}}e^{\left[ i\left( \omega -2\omega
_{a}\right) -\left( \kappa +\Gamma \right) \right] \frac{x}{2v_{c}}}+\left(
x_{2}\leftrightarrow x_{1}\right) \right] , \\
\phi _{oe}\left( x_{1},x_{2}\right) &=&\frac{1}{\sqrt{2}}\left[ \phi
_{o,k_{1}}\left( x_{1}\right) \phi _{e,k_{2}}\left( x_{2}\right) +\phi
_{e,k_{1}}\left( x_{2}\right) \phi _{o,k_{2}}\left( x_{1}\right) \right] , \\
\phi _{oo}\left( x_{1},x_{2}\right) &=&\frac{1}{\sqrt{2}}\left[ \phi
_{o,k_{1}}\left( x_{1}\right) \phi _{o,k_{2}}\left( x_{2}\right) +\phi
_{o,k_{1}}\left( x_{2}\right) \phi _{o,k_{2}}\left( x_{1}\right) \right] , \\
\phi _{ae}\left( x_{i}\right) &=&\theta \left( -x_{i}\right) \left( \mu
_{k_{1}}e^{ik_{1}x_{i}}+\mu _{k_{2}}e^{ik_{2}x_{i}}\right) +\theta \left(
x_{i}\right) \left( \eta _{k_{1}}e^{ik_{1}x_{i}}+\eta
_{k_{2}}e^{ik_{2}x_{i}}+\xi e^{\lambda _{-}x_{i}}\right) , \\
\phi _{oa}\left( x_{i}\right) &=&\rho _{k_{1}}e^{ik_{1}x_{i}}+\rho
_{k_{2}}e^{ik_{2}x_{i}}, \\
\phi _{aa} &=&-\frac{\overline{V}}{\sqrt{2}}\frac{\left( \mu _{k_{1}}+\mu
_{k_{2}}\right) }{\left( \Delta _{a}+U-i\frac{\kappa +\Gamma }{2}\right) },
\end{eqnarray}%
with
\begin{eqnarray}
\phi _{e,k_{i}}\left( x_{j}\right) &=&\frac{1}{\sqrt{2\pi }}\left[ \theta
\left( -x_{j}\right) +t_{k_{i}}\theta \left( x_{j}\right) \right]
e^{ik_{i}x_{j}}, \\
\phi _{o,k_{i}}\left( x_{j}\right) &=&\frac{1}{\sqrt{2\pi }}e^{ik_{i}x_{j}},
\end{eqnarray}%
where $\theta \left( x\right)$ is the step function; $\Gamma =\overline{V}^{2}/v_{c}$, $x_{c}=(x_{2}+x_{1})/2$ and $x=x_{2}-x_{1}$ are the
center-of-mass and the relative coordinate, respectively; $\Delta _{i}=v_{c}k_{i}-\omega /2$ and $\Delta
_{a}=\omega _{a}-\omega /2$;
\begin{eqnarray}
t_{k_{i}} &=&\frac{\Delta _{i}-\Delta _{a}+i\frac{\kappa -\Gamma }{2}}{%
\Delta _{i}-\Delta _{a}+i\frac{\kappa +\Gamma }{2}}, \\
\mu _{k_{1}} &=&\frac{1}{2\pi }\frac{\overline{V}}{\Delta _{2}-\Delta _{a}+i%
\frac{\kappa +\Gamma }{2}}, \\
\mu _{k_{2}} &=&\frac{1}{2\pi }\frac{\overline{V}}{\Delta _{1}-\Delta _{a}+i%
\frac{\kappa +\Gamma }{2}}, \\
B&=&\frac{\overline{V}}{iv_{c}\sqrt{2}}\xi , \label{eq:A30}\\
\xi &=&\frac{i 4\pi\overline{V} U}{v_{c} \left( \Delta _{a}+U-i\frac{%
\kappa +\Gamma }{2}\right) }\mu _{k_{1}}\mu _{k_{2}},
\end{eqnarray}%
and $\eta _{k_{i}}=\mu _{k_{i}}t_{k_{i}}$, $\rho _{k_{i}}=\mu _{k_{i}}/\sqrt{%
2}$.
\end{widetext}

\bibliographystyle{apsrev}
\bibliography{ref}

\end{document}